\documentclass[sigconf,nonacm,10pt]{acmart}

\usepackage{enumitem}
\usepackage{times}%
\usepackage{url}
\usepackage{hyperref}
\usepackage[most]{tcolorbox}
\usepackage{microtype}

\newcommand{\RED}[1]{  {\color{red}{#1}}}

\newcommand{\superscript}[1]{\ensuremath{^{\textrm{#1}}}}

\newcommand{\stitle}[1]{\bigskip\noindent\textbf{#1}} 

\definecolor{myboxcolor}{RGB}{218, 233, 247}
\definecolor{myboxcolor2}{HTML}{9EC6EA}
\definecolor{dark-gray}{gray}{0.2}

\tcolorboxenvironment{myitemize}{blanker,
	before skip=6pt,after skip=6pt,
	borderline west={3mm}{0pt}{red}}

\usepackage{tabularx}

\newcommand{\tlinebold}  {\specialrule{1 pt}{0pt}{1pt}}		 
\newcommand{\blinebold}  {\specialrule{1 pt}{1pt}{0pt}}		

 \hypersetup{
    colorlinks,%
    citecolor=black,%
    filecolor=black,%
    linkcolor=black,%
    urlcolor=black,%
     pdftitle={Visual Analytics Challenges and Trends in the Age of AI:	The BigVis Community Perspective},
     pdfauthor={Nikos Bikakis, Panos K. Chrysanthis,Guoliang Li,George Papastefanatos,Lingyun Yu},
    pdfsubject={Survey, Human-AI interaction, Visual AI, Human-Centered AI, Interactive AI, Explainable AI, LLM, Visual Interactive ML, Information Visualization, Big Data, Human-data interaction, Opportunities},
    pdfkeywords={Survey, Human-AI interaction, Visual AI, Human-Centered AI, Interactive AI, Explainable AI, LLM, Visual Interactive ML, Information Visualization, Big Data, Human-data interaction, Opportunities}
}

\begin{document}

\pagestyle{empty}

\title{Visual Analytics Challenges and Trends in the Age of AI:
	The BigVis Community Perspective}
\titlenote{\textbf{To appear in ACM SIGMOD Record 2025}}

 \author{Nikos Bikakis}
\authornote{Corresponding author.}
  \affiliation{
   \institution{Hellenic Mediterranean University \&}
      \institution{Archimedes/Athena RC}
   \country{Greece}
 }

 \author{Panos K. Chrysanthis}
  \affiliation{
   \institution{University of Pittsburgh}
   \country{USA}
 }

 \author{Guoliang Li}
 \affiliation{
   \institution{Tsinghua University}
   \country{China}
 }

 \author{George Papastefanatos}
 \affiliation{
   \institution{Athena RC}
   \country{Greece}
 }

 \author{Lingyun Yu}
 \affiliation{
   \institution{Xi’an Jiaotong-Liverpool University}
   \country{China}
   \vspace*{-6pt}
 }

\maketitle


\vspace{-6pt}
\section{Introduction}
This report provides insights into the challenges,  emerging topics, and opportunities related to  human‒data interaction and visual analytics in the AI era.

The BigVis 2024\footnote{\href{https://bigvis.imsi.athenarc.gr/bigvis2024}{ ~7th \textit{Intl.\ Workshop on Big Data Visual Exploration \& Analytics}}, in   conjunction with the
		\href{https://vldb.org/2024}{50th  \textit{Intl.\ Conf.\ on Very Large Databases}} (VLDB~2024), Guangzhou, China. More details about the BigVis workshops can be found in \cite{BikakisPCPAFFHLMSSS24}.}
	  organizing committee conducted a survey among experts in the field. They invited the Program Committee  members and the authors of accepted papers to share their views.
  \textit{Thirty-two scientists} from diverse research communities, including Databases, Information Visualization, and Human‒Computer Interaction, participated in the study. These scientists, representing both industry and academia, provided valuable insights into the current and future landscape of the field.

In this report, we analyze the survey responses and compare them to the findings of a similar study conducted four years ago \cite{AndrienkoADFFIK20}.
 The results reveal some interesting insights.
First, \textit{many of the critical challenges identified in the previous survey remain highly relevant today}, despite being unrelated to AI.
Meanwhile, the field’s landscape has significantly evolved, with \textit{most of today’s vital challenges not even being mentioned in the earlier survey}, underscoring the profound impact of AI-related advancements.

By summarizing the perspectives of the research community, this report aims to shed light on the key challenges, emerging trends, and potential research directions in human‒data interaction and visual analytics in the AI era.

%

 \section{Survey Overview}
 \label{sec:overview}
 The survey is divided into two parts. The first is related to challenges (Sec.~\ref{sec:challenges}), and the second focuses on emerging research topics (Sec.~\ref{sec:topics}).

 The participants were requested to answer six questions,  either by filling out free-text fields or selecting from the options provided.
 The survey was anonymous, since the questions related to personal information are optional, e.g., name, county, affiliation. The survey required, on average, about three to five minutes to be completed.

\stitle {Participants Demographics.}  We intended to find scientists from different
research communities (e.g.,~Databases, Information Visualization, HCI), and from industry and academia.
To this end, the survey was disseminated to the BigVis 2024 Program Committee members (58 members) and to the authors of accepted BigVis 2024 papers (34~authors).
 At the end, 32 \textit{of the scientists invited completed the survey}.

\begin{figure*}[t]
 	  	\centering
  	 \includegraphics[width=0.75\linewidth]{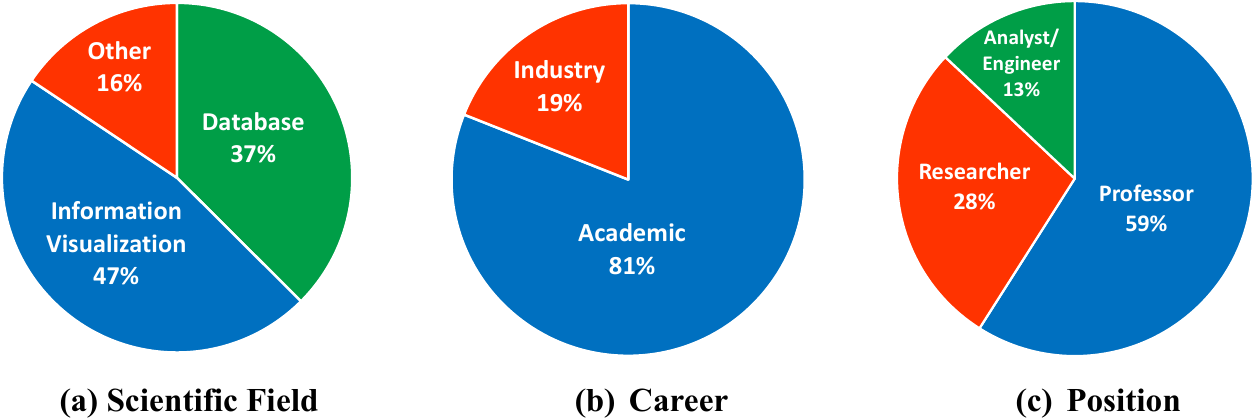}
 	  \caption{Participants Demographics}
 	  \label{fig:demogr}
 	   \vspace{8pt}
 	   \end{figure*}

 The following \textit{characteristics of the participants are collected} (Fig.~\ref{fig:demogr}):

 \begin{itemize}[leftmargin=7mm]

 	\item
 	\textbf{Scientific Field} (Fig.~\ref{fig:demogr}a):
 	The options were:
 	(a)~\textit{Database};
 	(b)~\textit{Information Visualization};
 	 (c)~\textit{Data Minning};
 	(d)~ \textit{Human‒Computer Interaction};
 	 (e)~\textit{Computer Graphics}; and
 	 (f)~\textit{Other}. Most of the participants belong to Information Visualization ($47$\%) and Database ($37$\%) communities, while $16$\% belong to others research fields.

\vspace{2pt}

 	\item
 	\textbf{Career} (Fig.~\ref{fig:demogr}b): The options were: \textit{Academic} ($81$\%) and \textit{Industry} ($19$\%).

\vspace{2pt}

 	\item
 	\textbf{Position} (Fig.~\ref{fig:demogr}c): The options were: \textit{Professor} ($59$\%); {\textit{Researcher} ($28$\%); and
 	\textit{Analyst/Scientist/ Engineer}~($13$\%)}.
 \end{itemize}

\begin{figure*}[t]
	\centering
	 	 \vspace*{6pt}
	\includegraphics[width=0.97\linewidth]{ 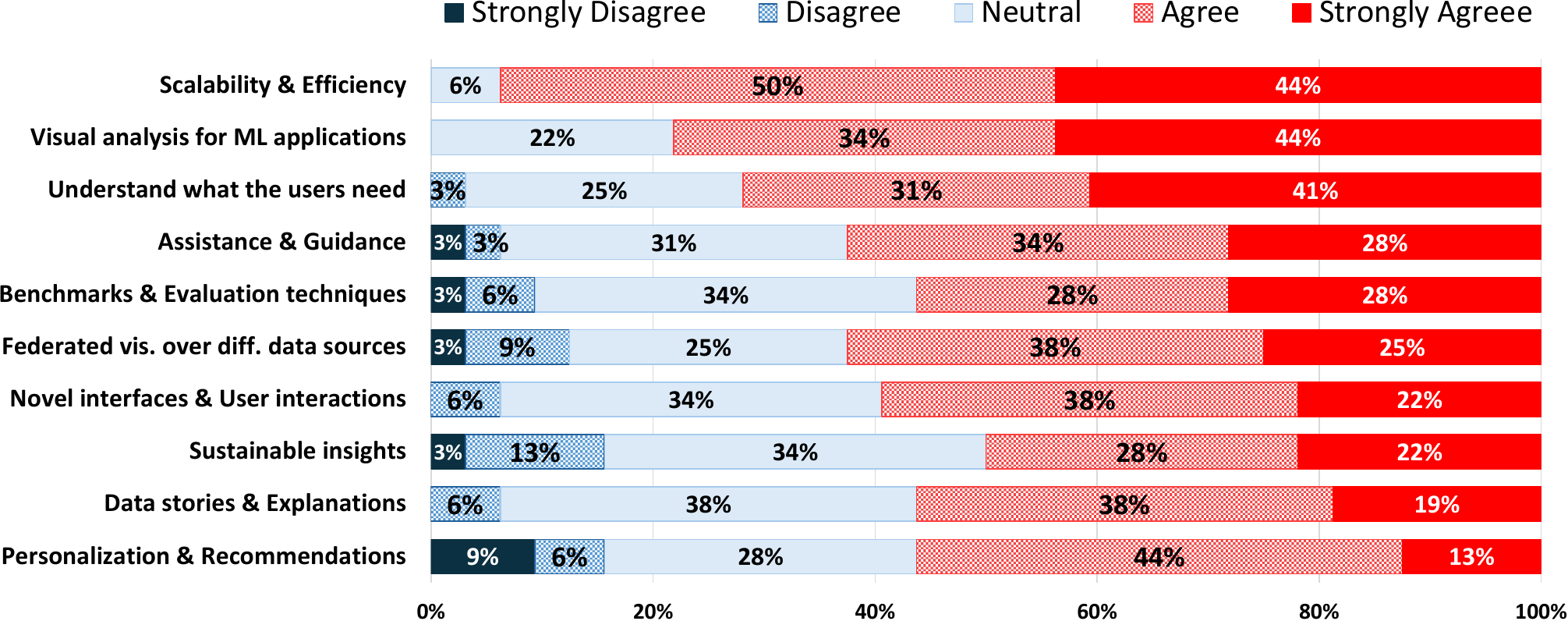}
	\caption{The Importance of the 2020 Challenges Today
		\textnormal{["\textit{Is this challenge important today}?"] }}
	\label{fig:chall2020}
	\vspace{6pt}
\end{figure*}

\section{Research Challenges}
\label{sec:challenges}

In this section we outline the survey’s results regarding research challenges related to data visualization and visual analytics.

In the first part (Sec.~\ref{sec:chall_20202}), the participants were asked to vote on today’s importance of the challenges emerged four years ago in a 2020 report, titled ``\textit{Big Data Visualization and Analytics: Future Research Challenges and Emerging Applications}'' \cite{AndrienkoADFFIK20}.
In the next part (Sec.~\ref{sec:chall_survey}), participants suggested a challenge they consider the most important today, regardless of whether it was included in 2020 challenges.

\subsection{The Challenges of the 2020 Report}
\label{sec:chall_20202}

This section presents the results of the survey regarding today’s importance of the ten challenges identified four years ago in the 2020 report. Particularly, in the context of the 3rd International Workshop on Big Data Visual Exploration and Analytics (BigVis 2020), the organizing committee invited 14 \textit{distinguished scientists}, from different communities to provide their insights regarding the \textit{challenges and the applications they find more interesting in coming years, related to the areas of Big Data visualization and analytics}.

The challenges indicated in the 2020 report were:
 (a)~\textit{Support scalability \& efficiency};
 (b)~\textit{Enable visual analysis for ML applications};
  (c)~\textit{Understand what the users need};
  (d)~\textit{Build novel interfaces \& user interactions};
  (e)~\textit{Assistance \& guidance};
  (f)~\textit{Generate data stories \& explanations};
   (g)~\textit{Enable federated visualization over different data sources};
    (h)~\textit{Develop benchmarks \& evaluation techniques};
    (i)~\textit{Provide sustainable insights}; and
    (k)~\textit{Enable personalization \& recommendations}.

\begin{tcolorbox}[tile, title={sharp corners},	title=Question 1, 	fonttitle=\bfseries,
	colback=myboxcolor, colbacktitle=myboxcolor2!80!black, coltitle=black, 
	top=3pt, bottom=3pt, left=3pt, right=3pt,
	toptitle=1pt,bottomtitle=1pt,lefttitle=3pt, 
]
	 The participants were asked to vote on the \textit{ten~challenges stated in the 2020 report, based on their importance/emerge}.
	 Particularly, the participants rated each challenge using a five-level Likert scale (i.e., Strongly Disagree, Disagree, Neutral, Agree, and Strongly Agree) on the question "\textit{Is this challenge important today}?".
	\end{tcolorbox}

\stitle{Question 1 Responses.} The results are presented in Figure~\ref{fig:chall2020} via a percent stacked bar chart.
Participants voted ``\textit{Scalability \& efficiency}'' as the most important challenge,
with $94$\% of participants indicating they agree or strongly agree, and $0$\%~disagree or strongly disagree. The second most significant is the ``\textit{Visual analysis for ML applications}'' challenge, with $78$\% (resp.\ $0$\%) of the participants state that agree or strongly agree (resp. disagree or strongly disagree). ``\textit{Sustainable insights}'' is the challenge in which most of the participants disagree ($13$\%) or strongly disagree~($3$\%). Finally, ``\textit{Personalization \& recommendations}'' is voted as the least important challenge.

It is worth mentioning that for \textit{all challenges, at least~$50$\% of the participants strongly agree or agree on today’s importance of the challenge. Similar results can be observed when considering importance scores}.

\subsection{The Challenges of the 2024 Survey}
\label{sec:chall_survey}
This section presents the challenges stated by the participants as the most important, regardless of whether they were included in 2020 challenges.

\begin{tcolorbox}[tile, title={sharp corners},	title=Question 2, 	fonttitle=\bfseries,
	colback=myboxcolor, colbacktitle=myboxcolor2!80!black, coltitle=black, 
	top=3pt, bottom=3pt, left=3pt, right=3pt,
	toptitle=1pt,bottomtitle=1pt,lefttitle=3pt, 
	]

	The participants were asked to provide in a free text the \textit{challenge they consider most important for the coming years}, along with a brief description.
\end{tcolorbox}

\noindent
\textbf{Question 2 Responses.}
The participants \textit{indicated} 16~\textit{challenges}. The challenges are presented in Table~\ref{tab:challenges}; the number in the parentheses that appears in some challenges \textit{indicates the number of participants that mentioned this challenge}.
Furthermore,
\textit{red font highlights the challenges that are mentioned for the first time in this survey}, i.e.,~challenges that were not mentioned in the 2020 survey.

The most commonly suggested challenge is the ``\textit{Use of LLMs in visualization and analytics}'' (voted by $16$\% of the participants),
whereas ``\textit{Fairness and Trustworthiness}'', as well as ``\textit{Visualization for non-expert users}'' are the next most common (each voted by $13$\%).
Note that none of the terms LLMs, fairness or trustworthiness is mentioned in the four years ago challenges.
Also note that \textit{explainability}, which emerged as one of the most frequently mentioned challenges, is also not mentioned in the 2020 report.

Other common challenges (mentioned by at least two participants) are related to: ``\textit{User assistance \& guidance}''; ``\textit{Understanding what the users need}'';  ``\textit{High dimensional \& stream data}''; ``\textit{Progressive data analysis \& visualization}''; and ``\textit{Immersive visualization}''. Among these challenges, ``\textit{High dimensional \& stream data}'', ``\textit{Progressive analysis}'', and ``\textit{Immersive visualization}'' appeared for the first time.

\begin{table}[t]
	\centering
	\caption{Survey Challenges $^\star$}
	\label{tab:challenges}
 	\small
	\vspace{-6pt}
	\begin{tabular}{>{\centering\arraybackslash}p{0.9\linewidth}}

		\tlinebold

		\RED{Exploit LLMs}$^{(5)}$, \RED{Ensure fairness \& trustworthiness}$^{(4)}$,
		\RED{Enable visualization for non-expert users}$^{(4)}$,  \linebreak
		Offer assistance \& guidance$^{(2)}$,
        Generate explanations$^{(2)}$,
        Understand what the users need$^{(2)}$,
        \RED{Handle high dimensional \& stream data}$^{(2)}$, \linebreak \RED{Enable progressive data analysis \& visualization}$^{(2)}$, \RED{Develop immersive visualization systems}$^{(2)}$,
        Provide sustainable insights,
        Support data abstraction,
        Implement novel scalable interfaces,
		Design context-specific visualizations,
        Formulate fundamental visualization problems,
        Use surrogate modeling,  \linebreak
        Develop energy consumption-based solutions
        \vspace{4pt}
		\\
		\blinebold
	\end{tabular}
	{
		\begin{flushleft}
			\hspace{0.4cm}
			{\footnotesize
				$^\star$ c\superscript{(x)}: x indicates the number of participants that mention the \\
				\hspace*{0.65cm} challenge c in the survey.   \RED{Red font: Challenges mentioned \\
					\hspace*{0.65cm} for the first time in the current survey (considering only \\
					\hspace*{0.65cm}  those indicated by at least two participants)}.}
	\end{flushleft}}
	\vspace*{-4pt}
\end{table}

\begin{figure*}[t]
	\centering
	\vspace*{-8pt}
	\includegraphics[width=0.94\linewidth]{ 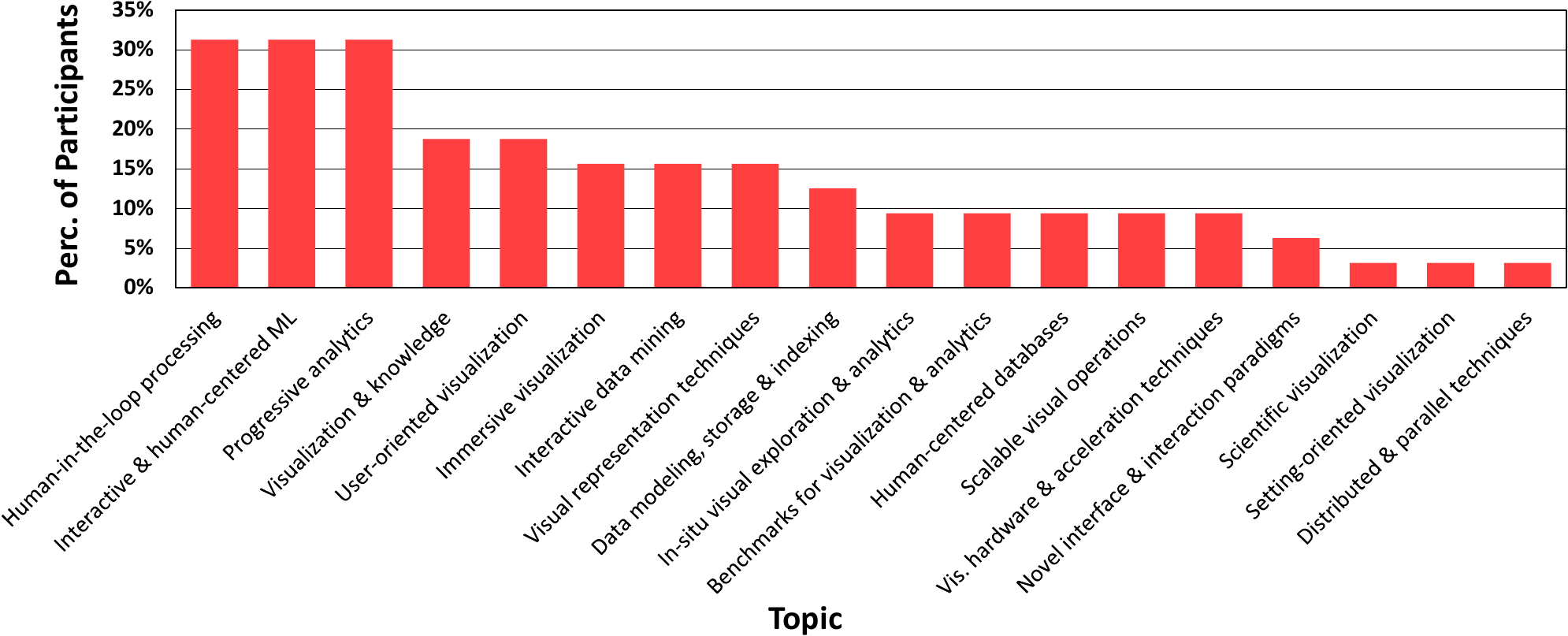}
			\vspace{-4pt}
	\caption{Emerging Topics: The Percentage of Participants that Voted for Each Topic}
	\label{fig:topic}
\end{figure*}

%

\section{Emerging Topics}
\label{sec:topics}

In this section we present the participants’ responses regarding the most emerging research topics in  Big data visualization and analytics field. The candidate list consists of the topics of interest included in the BigVis call-for-papers.

\begin{tcolorbox}[tile, title={sharp corners},	title=Question 3, 	fonttitle=\bfseries,
	colback=myboxcolor, colbacktitle=myboxcolor2!80!black, coltitle=black, 
	top=3pt, bottom=3pt, left=3pt, right=3pt,
	toptitle=1pt,bottomtitle=1pt,lefttitle=3pt, 
	]
	The participants were asked to\textit{ select (vote) from a list of candidate topics, up to three topics that they consider the most emerging}.
\end{tcolorbox}

\noindent
\textbf{Question 3 Responses.} Figure~\ref{fig:topic} shows the percentage
of the participants’ vote for each topic. The topics with the most votes are
``\textit{Human‒in‒the‒loop processing}'';
``\textit{Interactive \& human‒centered ML}''; and
``\textit{Progressive analytics}'', where $31$\% of the participants select. On the other hand, the topics with the less votes are:
``\textit{Scientific visualization}'';
``\textit{Setting-oriented visualization}''; and
``\textit{Distributed \& parallel techniques}'', which are voted~by~$3$\% of the participants.

%

\section{Discussion}
\label{sec:disc}

First, the survey \textit{highlights the broad acceptance of the importance of all the challenges identified four years ago} (Question~1), with at least half of the participants strongly agreeing or agreeing on their today’s importance.

The results regarding the current challenges (Question 2) revealed the importance of AI‒related problems. Notably, \textit{the most frequently mentioned challenges today were entirely absent four years ago}.
For example, problems related to LLMs, fairness \& trustworthiness, and explanations are some of the newcomers. Furthermore, challenges such as ``\textit{Non-expert users}'', ``\textit{High dimensional \& stream data}'', ``\textit{Progressive analysis}'', and  ``\textit{Immersive visualization}'' appeared also for the first time.

Further comparison of responses reveals that ``\textit{Scalability \& efficiency}'', the most important challenge in 2020 (Question~1), was not mentioned in Question~2. One possible explanation is that nearly $95$\% of participants had already rated it in Question 1 as (very) important, reducing the need to highlight it again. Similarly, ``\textit{Visual analysis over ML applications}'', the second most important challenge in 2020, was absent from Question~2 responses, despite being recognized as one of the most emerging topics (Question 3).

Additional discussion of the current challenges and
\mbox{state-of-the-art} approaches can be found in
\mbox{\cite{WuDCLKMMSVW23,blog,WangLZ24,AndrienkoAAWR22,YuanCYLXL21,Bikakis2022,BasoleM24,RaeesMLKP24,WuWSMCZZQ22,battle2020structured,YangLWL24,YeHHWXLZ24,WangCWQ22}}.

\bibliographystyle{ACM-Reference-Format}
\bibliography{biblio}

\end{document}